\documentclass[12pt]{article}
\usepackage{amssymb}

\begin{document}

\def\subsubsection{\@startsection{subsubsection}{3}{\z@}{-3.25ex plus
 -1ex minus -.2ex}{1.5ex plus .2ex}{\large\sc}}
\renewcommand{\thesection}{\arabic{section}}
\renewcommand{\thesubsection}{\arabic{section}.\arabic{subsection}}
\renewcommand{\thesubsubsection}
{\arabic{section}.\arabic{subsection}.\arabic{subsubsection}}
\renewcommand{\theequation}{\arabic{section}.\arabic{equation}}
\pagestyle{plain}
\sloppy
\textwidth 155mm
\textheight 620pt
\topmargin 35pt
\headheight 0pt
\headsep 0pt
\topskip 1pt
\oddsidemargin 0mm
\evensidemargin 10mm
\setlength{\jot}{4mm}
\setlength{\abovedisplayskip}{7mm}
\setlength{\belowdisplayskip}{7mm}
\newcommand{\be}{\begin{equation}}
\newcommand{\bel}[1]{\begin{equation}\label{#1}}
\newcommand{\ee}{\end{equation}}
\newcommand{\bea}{\begin{eqnarray}}
\newcommand{\ba}{\begin{array}}
\newcommand{\eea}{\end{eqnarray}}
\newcommand{\ea}{\end{array}}
\newcommand{\noin}{\noindent}
\newcommand{\ra}{\rightarrow}
\newcommand{\txs}{\textstyle}
\newcommand{\disp}{\displaystyle}
\newcommand{\scs}{\scriptstyle}
\newcommand{\scscs}{\scriptscriptstyle}
\newcommand{\sx}[1]{\sigma^{\, x}_{#1}}
\newcommand{\sy}[1]{\sigma^{\, y}_{#1}}
\newcommand{\sz}[1]{\sigma^{\, z}_{#1}}
\newcommand{\sP}[1]{\sigma^{\, +}_{#1}}
\newcommand{\sM}[1]{\sigma^{\, -}_{#1}}
\newcommand{\spm}[1]{\sigma^{\,\pm}_{#1}}
\newcommand{\Spm}{S^{\,\pm}}
\newcommand{\Sz}{S^{\, z}}
\newcommand{\hspf}{\hspace*{5mm}}
\newcommand{\hspt}{\hspace*{2mm}}
\newcommand{\vspf}{\vspace*{5mm}}
\newcommand{\vspt}{\vspace*{2mm}}
\newcommand{\hsix}{\hspace*{6mm}}
\newcommand{\hfour}{\hspace*{4mm}}
\newcommand{\vsix}{\vspace*{6mm}}
\newcommand{\vfour}{\vspace*{4mm}}
\newcommand{\vtwo}{\vspace*{2mm}}
\newcommand{\htwo}{\hspace*{2mm}}
\newcommand{\honecm}{\hspace*{1cm}}
\newcommand{\vonecm}{\vspace*{1cm}}
\newcommand{\htwocm}{\hspace*{2cm}}
\newcommand{\vtwocm}{\vspace*{2cm}}
\newcommand{\ru}{\rule[-2mm]{0mm}{8mm}}
\newcommand{\tinf}{\rightarrow \infty}
\newcommand{\udl}{\underline}
\newcommand{\ovl}{\overline}
\newcommand{\nwl}{\newline}
\newcommand{\nwp}{\newpage}
\newcommand{\clp}{\clearpage}
\newcommand{\simleq}{\raisebox{-1.0mm}
{\mbox{$\stackrel{\textstyle <}{\sim}$}}}
\newcommand{\simgeq}{\raisebox{-1.0mm}
{\mbox{$\stackrel{\textstyle >}{\sim}$}}}
\newcommand{\half}{\mbox{\small$\frac{1}{2}$}}
\newcommand{\smfrac}[2]{\mbox{\small$\frac{#1}{#2}$}}
\newcommand{\bra}[1]{\mbox{$\langle \, {#1}\, |$}}
\newcommand{\ket}[1]{\mbox{$| \, {#1}\, \rangle$}}
\newcommand{\exval}[1]{\mbox{$\langle \, {#1}\, \rangle$}}
\newcommand{\BIN}[2]
{\renewcommand{\arraystretch}{0.8}
\mbox{$\left(\ba{@{}c@{}}{\scs #1}\\{\scs #2}\ea\right)$}
\renewcommand{\arraystretch}{1}}
\def\gsim {\mbox{\hbox{ \lower-.6ex\hbox{$>$}
\kern-1.12em \lower.5ex\hbox{$\sim$}\kern+.35em}}}
\def\lsim {\mbox{\hbox{ \lower-.6ex\hbox{$<$}
\kern-1.12em \lower.5ex\hbox{$\sim$}\kern+.35em}}}

\begin{titlepage}
\thispagestyle{empty}
\begin{center}
\vspace*{1cm}
{\Large \bf
Exact Solution of the Master Equation for the Asymmetric Exclusion Process
}\\[25mm]

{\large {\sc
Gunter M. Sch\"utz}
} \\[8mm]

\begin{minipage}[t]{13cm}
\begin{center}
{\small\sl
Institut f\"ur Festk\"orperforschung,\\
Forschungszentrum J\"ulich, 52425 J\"ulich, Germany\\
e-mail: g.schuetz@kfa-juelich.de
}
\end{center}
\end{minipage}
\vspace{6cm}
\end{center}
\begin{center}
Abstract
\end{center}
{\small
Using the Bethe ansatz, we obtain the exact solution of the master equation
for the totally asymmetric exclusion process on an infinite one-dimensional
lattice. We derive explicit expressions for the
conditional probabilities $P(x_1,\dots,x_N;t|y_1,\dots,y_N;0)$ of finding $N$
particles on lattice sites $x_1,\dots,x_N$ at time $t$ with initial occupation
$y_1,\dots,y_N$ at time $t=0$.
}
\\
\vspace{1cm}\\
\udl{Key words:} Asymmetric exclusion process, Bethe ansatz\\
\udl{PACS numbers:} 05.40+j, 02.50Ga, 05.70.Ln
\end{titlepage}
\newpage

\newpage

\section{Introduction}
\setcounter{equation}{0}

Driven lattice gases and particularly the one-dimensional asymmetric simple
exclusion process (ASEP) have been intensively studied over the past decade
for a variety of reasons \cite{ZS}. The ASEP (for a definition see below) has
been suggested already in 1968 as a model for the kinetics of biopolymerization
\cite{1}. Two years later this process was introduced into the mathematical
literature \cite{Spit70} where it has received considerable attention
in the context of interacting particle systems \cite{Ligg85}. More
recently the
ASEP has been studied mainly by physicists as a model for polymers in
random media and as a dynamical model for interface growth \cite{KS}. It is
also
a discrete version of the noisy Burgers equation \cite{Burg74} and thus of
interest for the study of shocks \cite{Derr93,Ferr94} and for traffic models
\cite{Nage96}. While there are exact solutions and a good understanding of the
stationary behaviour of the system with
(trivial) periodic and (non-trivial) open boundary conditions with injection
and absorption of particles \cite{SD} - \cite{Essl96}, exact results for the
dynamics of the model are scarce (see e.g. Refs. \cite{Ligg85,Rost82} and, for
more recent work, Refs. \cite{Ferr94,Gwa92,DE} and references therein. It is
the
aim of this paper to present a new approach to this open problem by explicitly
solving the master equation for the system defined on an infinite lattice.

We study the ASEP with sequential updating. In the totally asymmetric version
of this very simple model each lattice site can be occupied by at most one
particle and particles hop with rate 1 to their right neighbouring site if this
was empty. If it was occupied, the attempted move is rejected. This defines a
Markov process with state space $X=\{0,1\}^S$ where a given configuration
$\udl{n} \in X$ is the set of occupation numbers $n_k = 0,1$ with the site
label $k \in S$. Alternatively, if one restricts oneselves to studying the
system with an arbitrary, but finite number of particles, one may define the
process on $Y = \{\{\emptyset\},
\{k_1\}, \{k_1,k_2\}, \dots,\}$ which is the collection of all finite subsets
of
$S$. In this case one identifies a configuration $\udl{n}$ which has particles
on sites $k_1, \dots, k_N$ with the corresponding set in $Y$. Here we consider
the system with finitely many particles $N$ defined on an infinite
one-dimensional lattice $S=\mathbb{Z}$.

A convenient presentation of the ASEP is in terms of a master equation for the
probability $P(B_N;t)$ of finding $N$ particles on sites $B_N = \{
k_1,\dots,k_N
\} \in Y$ at time $t$. Defining the ASEP alternatively in terms of a master
equation on the state space $X$ rather than $Y$ has turned out to be useful in
previous work. In this case the stochastic
time
evolution is manifestly seen to be generated by the quantum Hamiltonian
of a spin-1/2 Heisenberg chain. This suggests the use of the Bethe ansatz
and the quantum group symmetry for the calculation of energy gaps (which
give e.g. the dynamical exponent of the system) \cite{Gwa92,Kim95}
and certain time-dependent correlation functions \cite{Sand94b,Schu97}.
For the purposes of this paper using the coordinate
representation $Y$ of the state space is more transparent, as in this case
the master equation can be solved directly and explicitly using the coordinate
Bethe ansatz \cite{Beth31,Yang66}. While here we consider mainly the totally
asymmetric case, the partially asymmetric exclusion process may be solved
in the same way. We discuss this for the two-particle problem.

In fact, even though the two-particle system may seem trivial, it exhibits
already some of the characteristic behaviour of the driven system at finite
particle density. In the undriven system (i.e. in the symmetric
exclusion process) the diffusive spreading of a local perturbation does not
depend on the overall density. In contrast, in the asymmetric case the center
of mass of a local perturbation in an otherwise homogeneous finite background
density is, from numerical work, known to spread superdiffusively
\cite{ZS}. This implies a divergent collective diffusion constant as the number
of particles tends to infinity. There are, however, no quantitative
results for a system with a small number of particles.
As we will show here, in a system of two
particles this behaviour appears as an increase of the (collective) diffusion
constant which turns out to be the single-particle diffusion constant plus
a term proportional to the square of the asymmetry in the hopping
rates.

The paper is organized as follows: Sec. 2 gives the main result of the paper.
We formulate the master equation and present a (non-constructive) proof
that the expression derived from the Bethe ansatz in Sec. 3 is indeed
the solution. This is done because the proof is elementary and no reference
to the Bethe ansatz is necessary. A reader not interested in the derivation
of the solution may therefore skip Sec. 3 where the solution is constructed.
As a simple application,
Sections~4 and 5 focus on properties of the two-particle system.
In Sec. 4 we briefly study the partially asymmetric process where particles
are allowed to move both to the right and to the left, but with different
rates.
We restrict ourselves to the exact solution for the two-particle system, but
also explain how to obtain the solution for the general $N$-particle case.
In Sec. 5 we obtain a very simple new result, which shows that already in the
two-particle system the diffusive behaviour of the driven system is
substantially different from the symmetric, undriven process.
In Sec. 6 we present our conclusions.

\section{Solution of the master equation}
\setcounter{equation}{0}

Let $P(B_N;t)$ be the probability of finding $N$ particles on the set of sites
$B_N = \{k_1,\dots,k_N\}$ at time $t$. When considering the probability $P$ as
a function of the coordinates $k_i$ we always assume this set to be ordered,
$k_i < k_{i+1} \; \forall \; i$. It is important to note that as a function of
its arguments $k_i$ the function $P$ is well-defined in $\mathbb{Z}^N$, i.e.
also
for e.g. $k_i = k_{i+1}$ or $k_i > k_{i+1}$. However in this domain $P$ is not
a probability. In other words, in the domain ${\Omega}_N = k_1 < k_2 < \dots <
k_N \subset \mathbb{Z}^N$, the function $P$ is the probability defined above,
whereas in $\mathbb{Z}^N\setminus {\Omega}_N$ it is defined by the master
equation below, but is not a probability.

For the totally asymmetric exclusion process as described in the introduction,
$P(B_N;t)$ defined on $\mathbb{Z}^N\times [0,\infty)$ satisfies the master
equation
\bel{2-1}
\frac{d}{dt} P(k_1,\dots,k_N;t) = P(k_1-1,\dots,k_N;t) + \dots +
P(k_1,\dots,k_N -1 ;t) - N P(k_1,\dots,k_N;t) .
\ee
This has to be supplemented by boundary conditions in $\mathbb{Z}^N$. If any
two
neighbouring arguments $k_i,k_{i+1}$  are equal, $P$ satisfies
\bel{2-2}
P(k_1,\dots,k_i,k_{i+1}=k_i,\dots,k_N;t) =
P(k_1,\dots,k_i,k_{i+1}=k_i+1,\dots,k_N;t) \; \forall \; t \geq 0 .
\ee
This boundary condition expresses the exclusion interaction. This is easy
to see in the simplest case of two particles. Then (\ref{2-1}) and (\ref{2-2})
defined on $\mathbb{Z}^2$ read
\bel{2-3}
\frac{d}{dt} P(k_1,k_2;t) = P(k_1-1,k_2;t) +
P(k_1,k_2 -1 ;t) - 2 P(k_1,k_2;t)
\ee
and
\bel{2-4}
P(k,k;t)  =  P(k,k+1;t) \honecm \forall \; k \mbox{ and }
t \geq 0
\ee
which is equivalent to the following equations with $P$ restricted to
${\Omega}_2$
\bel{2-5}
\frac{d}{dt} P(k_1,k_2;t) = P(k_1-1,k_2;t) +
P(k_1,k_2 -1 ;t) - 2 P(k_1,k_2;t) \mbox{ if } k_2 - k_1 > 1
\ee
and
\bel{2-6}
\frac{d}{dt} P(k_1,k_2;t) = P(k_1 - 1 , k_2 ;t) -  P(k_1,k_2;t)
\mbox{ if } k_2 - k_1 = 1 .
\ee
The second equation (\ref{2-6}) expresses that due to exclusion the
configuration $(k,k+1)$ can be reached in a single step only from the
configuration $k-1,k+1$ and be left only in a single way (which is by moving
to $(k,k+2)$). Extending the range of validity of (\ref{2-5}) to all
$\mathbb{Z}^2$ requires adding
$P(k_1  , k_2-1 ;t) -  P(k_1,k_2;t)$ to (\ref{2-6}). However, by demanding that
(\ref{2-4}) holds for all times, this is equivalent to adding $0$. Thus
(\ref{2-5}) and (\ref{2-6}) remain unchanged, i.e., the two sets of equations
have the same solutions in the ``physical'' domain $\Omega_2$. It may seem more
natural to use the second formulation of the master equation which is in the
$N$-particle case the restriction of the validity of the master equation
(\ref{2-1}) to the domain ${\Omega}_N$. One could then obtain a set of
equations
equivalent to (\ref{2-1}), (\ref{2-2}) by replacing the boundary condition
(\ref{2-2}) by including appropriately chosen Kronecker delta-functions in
(\ref{2-1}). However, it turns out that solving the equation is more
straightforward and transparent in the formulation (\ref{2-1}), (\ref{2-2}).

We finally note that
with specified initial condition $A_N \equiv \{l_1,\dots,l_N\} \in Y$, i.e.,
\bel{2-2a}
P(B_N;0) = \delta_{A_N,B_N}
\ee
the probability $P(B_N;t)$ becomes
the conditional probability $P(B_N;t|A_N;0)$ and thus a complete
solution of the problem. Also as a function of the arguments $l_i$ the function
$P$ is a probability only in the domain $l_1 < \dots < l_N$.

Let us now introduce the function
\bel{2-7}
F_p(n;t) \equiv e^{-t} \sum_{k=0}^{\infty}
\BIN{k+p-1}{p-1} \frac{t^{k+n}}{(k+n)!}
\ee
where the binomial coefficient and the factorial are defined by the
$\Gamma$-function, i.e. $a! \equiv \Gamma(a+1)$ and
\bel{2-8}
\BIN{a}{b} \equiv \frac{\Gamma(a+1)}{\Gamma(b+1)\Gamma(a-b+1)}.
\ee
In what follows we shall need only $p,n \in Z$ and
$t \in [0,\infty)$. We list some of the properties of $F_p(n;t)$:\\

\noin (1) For integer $p \leq 0$, $F_p$ reduces to a finite sum,
\bel{2-10}
F_p(n;t) = e^{-t} \sum_{k=0}^{|p|} (-1)^k \BIN{|p|}{k} \frac{t^{k+n}}{(k+n)!}.
\ee
In particular,
\bel{2-12}
F_0(n;t) = \frac{t^{n}}{n!} e^{-t}
\ee\\

\noin (2) For the time derivative one finds
\bel{2-13}
\frac{d}{dt} F_p(n;t) = F_{p-1}(n-1;t) = F_p(n-1;t) - F_p(n;t),
\ee\\

\noin (3) and for the time integral one gets
\bel{2-14}
\int_{0}^t dt F_p(n;t) = F_{p+1}(n+1;t) - \BIN{-n-1+p}{p} =
\sum_{k=n+1}^{\infty} F_p(k;t) - \BIN{-n-1+p}{p}
\ee\\

\noin (4) At time $t=0$ one has
\bel{2-15}
\lim_{t \rightarrow 0} F_p(n;t) = \BIN{-n+p-1}{p-1}
\ee
which vanishes for $n > 0$.
Now we state the main result of this paper:\\

{\bf Theorem:} Let $F(B_N,A_N;t)$ be the $N\times N$ matrix with matrix
elements $F_{ij} = F_{i-j}(k_i-l_j;t)$.
Then
\bel{2-17}
P(B_N;t|A_N;0) = \det{F(B_N,A_N;t)}
\ee
is the solution of the master equation (\ref{2-1}) with boundary condition
(\ref{2-2}) and with initial condition (\ref{2-2a}). \\

The theorem states that the conditional probability of finding $N$ particles at
time $t$ on $B_N \subset \Omega_N$ if initially (at time $t=0$) they had been
on sites $A_N \subset \Omega_N$ is given by the determinant (\ref{2-17}).
How this result was derived is explained in the next section. Here we give
a proof of the theorem which is independent of this construction.
First we show that the determinant is a solution to
the master equation (\ref{2-1}). Then we show that it satisfies the boundary
condition (\ref{2-2}) and finally we prove that it satisfies the
initial condition (\ref{2-2a}).

\noin {\em Proof:} (i) Because of the factor $e^{-t}$ in the functions
$F_p(n;t)$ the matrix $F$ may be written $e^{-t} \tilde{F}$ and one gets
\be
\frac{d}{dt} \det{F} = e^{-Nt} \frac{d}{dt} \det{\tilde{F}} - N \det{F}.
\ee
This accounts for the term $-N P$ on
the r.h.s. of (\ref{2-1}). The time derivative of the determinant of
$\tilde{F}$
may be written
\bea
\frac{d}{dt}
\left| \ba{llll}
\tilde{F}_{11} & \tilde{F}_{12} & \dots  & \tilde{F}_{1N} \\
\tilde{F}_{21} & \tilde{F}_{22} & \dots  & \tilde{F}_{2N} \\
\vdots         & \vdots         & \vdots & \vdots         \\
\tilde{F}_{N1} & \tilde{F}_{N2} & \dots  & \tilde{F}_{NN}
\ea \right|
 & = &
\left| \ba{llll}
\dot{\tilde{F}}_{11} & \dot{\tilde{F}}_{12} & \dots  & \dot{\tilde{F}}_{1N} \\
\tilde{F}_{21} & \tilde{F}_{22} & \dots  & \tilde{F}_{2N} \\
\vdots         & \vdots         & \vdots & \vdots         \\
\tilde{F}_{N1} & \tilde{F}_{N2} & \dots  & \tilde{F}_{NN}
\ea \right| \nonumber \\
 &   & +
\left| \ba{llll}
\tilde{F}_{11} & \tilde{F}_{12} & \dots  & \tilde{F}_{1N} \\
\dot{\tilde{F}}_{21} & \dot{\tilde{F}}_{22} & \dots  & \dot{\tilde{F}}_{2N} \\
\vdots         & \vdots         & \vdots & \vdots         \\
\tilde{F}_{N1} & \tilde{F}_{N2} & \dots  & \tilde{F}_{NN}
\ea \right| \nonumber \\
 &   & + \dots + \nonumber\\
 &   & +
\left| \ba{llll}
\tilde{F}_{11} & \tilde{F}_{12} & \dots  & \tilde{F}_{1N} \\
\tilde{F}_{21} & \tilde{F}_{22} & \dots  & \tilde{F}_{2N} \\
\vdots         & \vdots         & \vdots & \vdots         \\
\dot{\tilde{F}}_{N1} & \dot{\tilde{F}}_{N2} & \dots  & \dot{\tilde{F}}_{NN} \\
\ea \right| \nonumber
\eea
The matrix elements in row $i$ are $(\tilde{F}_{i-1}(k_i-l_1),
\tilde{F}_{i-2}(k_i-l_2),\dots,\tilde{F}_{i-N}(k_i-l_N))$ and their time
derivatives are $(\tilde{F}_{i-1}(k_i-1-l_1),
\tilde{F}_{i-2}(k_i-1-l_2),\dots,\tilde{F}_{i-N}(k_i-1-l_N))$, see
(\ref{2-13}). Thus each determinant in the time derivative of $\tilde{F}$
contributes exactly one of the terms $P(k_1,\dots,k_i-1,\dots,k_N;t)$ on the
r.h.s. of (\ref{2-1}), i.e. (\ref{2-17}) satisfies the master equation
(\ref{2-1}).

(ii) In order to show that (\ref{2-17}) satisfies the boundary condition
(\ref{2-2}) we note that according to (\ref{2-13}) each column in $F$ may be
written
\bea
\left(F_{i-1}(k_i-l_1),F_{i-2}(k_i-l_2),\dots,F_{i-N}(k_i-l_N)\right)
& & \nonumber \\
\hsix = \htwo
\left(F_{i-1}(k_i+1-l_1),F_{i-2}(k_i+1-l_2),\dots,F_{i-N}(k_i+1-l_N)\right)
& & \nonumber \\
\honecm +
\left(F_{i}(k_i+1-l_1),F_{i-1}(k_i+1-l_2),\dots,F_{i+1-N}(k_i+1-l_N)\right)
\eea
Assume now that $k_{i+1} = k_i+1$. This gives for the r.h.s. of (\ref{2-2})
\bea
\left| \ba{llll}
F_0(k_1-l_1;t) & F_{-1}(k_1-l_2;t)  & \dots  & F_{1-N}(k_1-l_N;t)  \\
\vdots         & \vdots         & \vdots & \vdots         \\
F_{i-1}(k_i-l_1;t) & F_{i-2}(k_i-l_2;t)  & \dots  & F_{i-N}(k_i-l_N;t)  \\
F_{i}(k_{i+1}-l_1;t) & F_{i-1}(k_{i+1}-l_2;t)  & \dots  & F_{i+1-N}
(k_{i+1}-l_N;t)  \\
\vdots         & \vdots         & \vdots & \vdots         \\
F_{N-1}(k_N-l_1;t) & F_{N-2}(k_N-l_2;t)  & \dots  & F_{0}(k_N-l_N;t)
\ea \right|
 &  & \nonumber \\
\hsix = \htwo
\left| \ba{llll}
F_0(k_1-l_1;t) & F_{-1}(k_1-l_2;t)  & \dots  & F_{1-N}(k_1-l_N;t)  \\
\vdots         & \vdots         & \vdots & \vdots         \\
F_{i-1}(k_i+1-l_1;t) & F_{i-2}(k_i+1-l_2;t)  & \dots  & F_{i-N}(k_i+1-l_N;t) \\
F_{i}(k_{i+1}-l_1;t) & F_{i-1}(k_{i+1}-l_2;t)  & \dots  & F_{i+1-N}
(k_{i+1}-l_N;t)  \\
\vdots         & \vdots         & \vdots & \vdots         \\
F_{N-1}(k_N-l_1;t) & F_{N-2}(k_N-l_2;t)  & \dots  & F_{0}(k_N-l_N;t)
\ea \right|
 &   & \nonumber \\
\honecm +
\left| \ba{llll}
F_0(k_1-l_1;t) & F_{-1}(k_1-l_2;t)  & \dots  & F_{1-N}(k_1-l_N;t)  \\
\vdots         & \vdots         & \vdots & \vdots         \\
F_{i}(k_i+1-l_1;t) & F_{i-1}(k_i+1-l_2;t)  & \dots  & F_{i+1-N}(k_i+1-l_N;t)
 \\
F_{i}(k_{i+1}-l_1;t) & F_{i-1}(k_{i+1}-l_2;t)  & \dots  & F_{i+1-N}
(k_{i+1}-l_N;t)  \\
\vdots         & \vdots         & \vdots & \vdots         \\
F_{N-1}(k_N-l_1;t) & F_{N-2}(k_N-l_2;t)  & \dots  & F_{0}(k_N-l_N;t)
\ea \right|
 &  & \nonumber \\
\hsix = \htwo
\left| \ba{llll}
F_0(k_1-l_1;t) & F_{-1}(k_1-l_2;t)  & \dots  & F_{1-N}(k_1-l_N;t)  \\
\vdots         & \vdots         & \vdots & \vdots         \\
F_{i-1}(k_i+1-l_1;t) & F_{i-2}(k_i+1-l_2;t)  & \dots  & F_{i-N}(k_i+1-l_N;t) \\
F_{i}(k_{i+1}-l_1;t) & F_{i-1}(k_{i+1}-l_2;t)  & \dots  & F_{i+1-N}
(k_{i+1}-l_N;t)  \\
\vdots         & \vdots         & \vdots & \vdots         \\
F_{N-1}(k_N-l_1;t) & F_{N-2}(k_N-l_2;t)  & \dots  & F_{0}(k_N-l_N;t)
\ea \right|
\eea
The second determinant in the sum on the r.h.s. of this equation vanishes
since two rows are identical. It remains the first determinant in the sum
which is equal to the l.h.s. of (\ref{2-2}).

(iii) It remains to show that (\ref{2-17}) satisfies the correct initial
condition (\ref{2-2a}). We first assume that $k_1 > l_1$. This implies
$k_i > l_1$ since $k_1 < \dots < k_N$. At $t=0$, all the matrix elements
$F_{i1}=F_{i-1}(k_i-l_1;0)$ vanish, see (\ref{2-15}), and hence the
determinant vanishes. Now we assume $k_1=l_1$. Then in the first column
$F_{11}=1$, but the other matrix elements $F_{i1}$ in the first column are
still
zero, since all $k_i$ are larger than $l_1$ except $k_1$. The
determinant of $F$ is therefore equal to the determinant of the matrix
$F^{(1)}$ obtained from $F$ by omitting the first row and the first column.

In the next step assume now first that $k_2 > l_2$. Repeating the previous step
gives $\det{F}(B_N,A_N;0)=\delta_{k_1,l_1}\delta_{k_2,l_2}\det{F^{(2)}}$.
Iterating this procedure $N$ times finally gives
$\det{F}(B_N,A_N;0) = \delta_{A_N,B_N}$.\hfill $\square$

An integral representation of the terms in the determinant is derived in the
next section (\ref{3-9}).

\section{Bethe ansatz solution}
\setcounter{equation}{0}

For a derivation of the solution the master equation (\ref{2-1}), (\ref{2-2}),
(\ref{2-2a}), we first turn the differential equations (\ref{2-1}) into an
eigenvalue problem by the ansatz $P(B_N;t) = e^{-\epsilon t} P(B_N)$. In order
to solve for the resulting difference equation we follow the strategy employed
by Bethe for the solution of the isotropic Heisenberg spin chain \cite{Beth31}
and extended by Yang and Yang \cite{Yang66} to the anisotropic spin
chain. Rather than using $k_i,l_i$ for the integer coordinates of the particles
we shall use in this section the notation $x_i,y_i \in Z$. For momentum labels
we shall use $p_i$.

First we consider $N=1$. The resulting equation is
\bel{3-1}
\epsilon P(x) = - P(x-1) + P(x)
\ee
which is readily solved by $P(x) = e^{ipx}$ with $p \in [0,2\pi )$. This gives
for the ``energy''
\bel{3-2}
\epsilon_p = 1 - e^{-ip}
\ee
and $P(x;t) = \int_0^{2\pi} dp e^{-\epsilon_pt} f(p) e^{ipx}$. The initial
condition $P(x;0) = \delta_{x,y}$ determines $f(p)=e^{-ipy}/(2\pi)$ and finally
yields
\bea
\label{3-3a}
P(x;t|y;0) & = & \frac{1}{2\pi} \int_0^{2\pi} dp e^{-\epsilon_p t - ipy}
                 e^{ipx} \\
\label{3-3b}
           & = & \frac{t^{x-y}}{(x-y)!} e^{-t} \\
\label{3-3c}
           & = & F_0(x-y;t)
\eea

For $N=2$ one has to solve
\bea
\label{3-4a}
\epsilon P(x_1,x_2) & = & - P(x_1-1,x_2) - P(x_1,x_2-1) + 2 P(x_1,x_2) \\
\label{3-4b}
P(x,x) & = & P(x,x+1) \;\;\; \forall \; x
\eea
The first equation is solved by Bethe's ansatz
\bel{3-5}
P(x_1,x_2) = A_{12} e^{ip_1x_1 + ip_2x_2} + A_{21}
              e^{ip_2x_1 + ip_1x_2}
\ee
with arbitrary constants $A_{ij}(p_1,p_2)$ and gives
\bel{3-6}
\epsilon_{p_1,p_2} = \epsilon_{p_1} + \epsilon_{p_2}.
\ee
The second equation (\ref{3-4b}) fixes the ratio $S_{12} \equiv A_{12}/A_{21}$.
Inserting (\ref{3-5}) gives
\bel{3-7}
S_{12} = - \frac{1-e^{ip_1}}{1-e^{ip_2}}
\ee

The range of values $p_1$ and $p_2$ may take needs some discussion.
In the usual Heisenberg quantum chain one finds a bound state in the
two-particle sector, i.e. a state with complex momenta $p_{1,2} = u \pm iv$.
This is a solution for vanishing $A_{12}$ or vanishing $A_{21}$ in which case
the wave function decays exponentially in the distance $x_2 - x_1$
(see next Section).  Here
there is no non-zero $p$ for which either $A_{12}$ or $A_{21}$ vanish and hence
no bound state. We conclude that $p_1,p_2 \in [0,2\pi)$ and
$P(x_1,x_2;t) = \int dp_1 \int dp_2 e^{-(\epsilon_{p_1} + \epsilon_{p_2})t}
f(p_1,p_2)( e^{ip_1x_1 + ip_2x_2} + S_{21} e^{ip_2x_1 + ip_1x_2})$ is the
general solution of (\ref{2-1}) with boundary condition (\ref{2-2}).
(For obvious reasons we define $S_{21} \equiv S_{12}^{-1}$.)

In order to satisfy the initial condition (\ref{2-2a}) one has to determine
$f(p_1,p_2)$ and discuss the pole resulting from the integration over $S_{21}$.
Assuming that the particles were initially at sites $y_1,y_2$
it turns out that choosing $f(p_1,p_2)=e^{-ip_1 y_1 - ip_2 y_2}$ and
defining the position of the pole in $S_{21}$ by $p_1 \rightarrow p_1 + i0$
gives the correct initial condition $P(x_1,x_2;0) =
\delta_{x_1,y_1} \delta_{x_2,y_2}$. This gives
\bea
\label{3-8a}
P(x_1,x_2;t|y_1,y_2;0) & = & \frac{1}{(2\pi)^2} \int_0^{2\pi} dp_1
\int_0^{2\pi} dp_2 e^{-(\epsilon_{p_1} + \epsilon_{p_2})t - ip_1 y_1 - ip_2
y_2}
 \nonumber \\
 & & \left( e^{ip_1x_1 + ip_2x_2} - \frac{1-e^{ip_2}}{1-e^{ip_1}}
     e^{ip_2x_1 + ip_1x_2}\right) \\
\label{3-8b}
           & = & F_0(x_1-y_1;t)F_0(x_2-y_2;t) \nonumber \\
           &   & - F_{-1}(x_1-y_2;t)F_1(x_2-y_1;t)\\
           & = & \det{F(B_2,A_2;t)}
\eea
with $B_2 = \{x_1,x_2\} \subset \Omega_2$, $A_2 = \{y_1,y_2\} \subset \Omega_2$
and the position of the pole in (\ref{3-8a}) defined as discussed above.

In the same way one continues for $N\geq 3$. One constructs a superposition
$\Psi=\sum A_{i_1,\dots,i_N} \exp{(i p_{i_1} x_1 + \dots + i p_{i_N} x_N)}$ of
$N$-particle plane waves with all $N!$ possible permutations of the wave
numbers
$p_i$ and with amplitudes $A_{i_1,\dots,i_N}$. The ratio of any two amplitudes
for plane waves where two momenta $p_i,p_j$ are interchanged is $S_{ij}$, in
the
same way as in the two-particle case (\ref{3-7}). This takes care of the
boundary condition
(\ref{2-2}) when (any) two particles are on nearest neighbour sites. The
crucial
point is that for higher number of particles there are no new constraints from
the boundary condition when more than two particles are on adjacent sites. This
can be seen by noting that satisfying the boundary condition for any given pair
is independent of the coordinates of the remaining particles. So one constructs
the Bethe wave function by starting from $e^{i p_1 x_1 + \dots + i p_Nx_N}$
with
amplitude $A_{12\dots N}=1$ and
then performing all possible permutations of the momenta. For each permutation
$(i,j)\rightarrow (j,i)$ one multiplies with a factor $S_{ji}$ as in
(\ref{3-8a}). The total ``energy'' $\epsilon$ corresponding to such a
wave function is the sum of the single particle energies
$\epsilon_{p_1,\dots,p_N} = \sum_{i=1}^N \epsilon_{p_i}$. The initial
condition (\ref{2-2a}) determines the overall normalization of the wave
function and the position of the poles arising from the integration over
the various $S_{ij}$ appearing in the wave function. It is satisfied by the
choice $f(p_1,\dots,p_N) = e^{-(ip_1y_1 + \dots ip_Ny_N)}$ and by placing
the poles in the same way as in the two-particle case, i.e., by setting
$p_i \rightarrow p_i +i0$ in the denominators.

This construction provides an integral representation of the $N!$ terms
appearing in the determinant (\ref{2-17}). Therefore the solution of the master
equation may be written
\bea
\label{3-9}
P(x_1,\dots,x_N;t|x_1,\dots,x_N;0) & = &
\prod_{j=1}^N \frac{1}{2\pi} \int_0^{2\pi} dp_j
e^{-\epsilon_{p_j}t - ip_jy_j}
\times \nonumber \\
 & & \Psi_{p_1,\dots,p_N}(x_1,\dots,x_N)
\eea
with the Bethe wave function $\Psi$ as defined above. For three particles
it reads
\bea
\label{3-10}
\Psi_{p_1,p_2,p_3}(x_1,x_2,x_3) & = & e^{ip_1x_1 + ip_2x_2 + ip_3x_3}
     + S_{21} e^{ip_2x_1 + ip_1x_2 + ip_3x_3} \nonumber \\
 & & + S_{21}S_{31} e^{ip_2x_1 + ip_3x_2 + ip_1x_3}
     + S_{21}S_{31}S_{32} e^{ip_3x_1 + ip_2x_2 + ip_1x_3} \nonumber \\
 & & + S_{31}S_{32} e^{ip_3x_1 + ip_1x_2 + ip_2x_3}
     + S_{32} e^{ip_1x_1 + ip_3x_2 + ip_2x_3}.
\eea

\section{The partially asymmetric process}
\setcounter{equation}{0}

Using the Bethe ansatz one may also solve for the partially asymmetric
process where particles are allowed to move with rate $D_L$ to left
and with rate $D_R$ to the right. The strategy is the same as for the
totally asymmetric case discussed in the preceding section. The main
difference in the analysis is the occurrence of bound states in addition to the
continuum. We shall discuss in some detail
only the one- and two-particle systems. This is, in principle, sufficient
to construct the general $N$-particle solution.

The case of a single particle can be copied with little modification
from the previous section. It is convenient to introduce the asymmetry
$q=e^a$ and the time scale $D$ by
\bea
\label{4-0a}
q & = & \sqrt{\frac{D_R}{D_L}} \\
\label{4-0b}
a & = & \ln{q} \\
\label{4-0c}
D & = & \sqrt{D_R D_L} .
\eea
After separating the time dependence the master equation reads
\bel{4-1}
\epsilon P(x) = - D_R P(x-1) - D_L P(x+1) + (D_L + D_R) P(x)
\ee
which is readily solved by $P(x) = e^{ipx}$ with $p \in [0,2\pi )$. This gives
for the ``energy''
\bel{4-2}
\epsilon_p = D_R(1 - e^{-ip}) + D_L(1 - e^{ip})
\ee
and $P(x;t) = \int_0^{2\pi} dp e^{-\epsilon_p} f(p) e^{ipx}$. The initial
condition $P(x;0) = \delta_{x,y}$ determines $f(p)=e^{-ipy}/(2\pi)$ and finally
yields
\bea
\label{4-3a}
P(x;t|y;0) & = & \frac{1}{2\pi} \int_0^{2\pi} dp e^{-\epsilon_p t - ipy}
                 e^{ipx} \\
\label{4-3b}
           & = & e^{-(q+q^{-1})Dt} q^{x-y}
                 I_{x-y}(2Dt)
\eea
where $I_n(\tau)$ is the modified Bessel function. The representation of
(\ref{4-3a}) in terms of the Bessel function (\ref{4-3b}) is obtained by
an elementary contour integration. It is easy to verify that both expressions
(\ref{4-3a}) and (\ref{4-3b}) satisfy the same differential-difference equation
with the same initial condition $P(x;0|y;0) = \delta_{x,y}$.

Note that the Bessel function diverges asymptotically $\sim e^{2Dt}$
which is not sufficiently fast to cancel the prefactor $e^{-(q+q^{-1})Dt}$
in (\ref{4-3b}). The interpretation of this observation is that the probability
of finding the particle at site $x$  with $x$ kept fixed decays exponentially
with
an inverse correlation time or ``energy gap'' $\Delta \epsilon =
D(q+q^{-1}-2)$.
However, by going into a comoving frame with velocity $v=D(q-q^{-1})=D_R-D_L$
(see Sec.~5), i.e. by studying the behaviour of the distribution around
$x' = x+vt$, one finds the usual algebraic, diffusive behaviour.

For $N=2$ one has to solve
\bea
\label{4-4a}
\epsilon P(x_1,x_2) & = & - D_R \left( P(x_1-1,x_2) + P(x_1,x_2-1)
                          - 2 P(x_1,x_2) \right) \nonumber \\
                    &   & - D_L \left( P(x_1+1,x_2) + P(x_1,x_2+1)
                          - 2 P(x_1,x_2) \right) \\
\label{4-4b}
P(x,x+1) & = & \frac{D_R P(x,x) + D_L P(x+1,x+1)}{D_R+D_L} \;\;\; \forall \; x
\eea
The first equation is solved by Bethe's ansatz
\bel{4-5}
P(x_1,x_2) = A_{12} e^{ip_1x_1 + ip_2x_2} + A_{21}
              e^{ip_2x_1 + ip_1x_2}
\ee
with arbitrary constants $A_{ij}(p_1,p_2)$ and gives for the two-particle
energy
\bel{4-6}
\epsilon_{p_1,p_2} = \epsilon_{p_1} + \epsilon_{p_2}.
\ee
The second equation (\ref{4-4b}) fixes the ratio $S_{12} \equiv A_{12}/A_{21}$.
Inserting (\ref{4-5}) gives
\bel{4-7}
S_{12} = - \frac{D_R + D_L e^{ip_1+ip_2} - (D_R+D_L)e^{ip_1}}
                {D_R + D_L e^{ip_1+ip_2} - (D_R+D_L)e^{ip_2}}
\ee
which depends only on the momenta $p_{1,2}$ and the asymmetry $q$.

As discussed in the preceding section here one finds besides the
continuum $p_1,p_2 \in [0,2\pi)$ a solution corresponding
to a bound state. To see this, we set $p_1=u - i(a-v)$, $p_2=u - i(a+v)$
with $u,v$ real and $a=\ln{q}$. Clearly, in order to obtain a wave
function which decays exponentially in $x_2-x_1>0$, either $A_{12}$ or $A_{21}$
must vanish. Choosing
\bel{4-8}
e^{-v} = \frac{\cos{u}}{\cosh{a}} < 1
\ee
gives $A_{12}\equiv D_R + D_L e^{ip_1+ip_2} - (D_R+D_L)e^{ip_1} = 0$. The
wave function (\ref{4-5}) reduces then to a single expression
\bel{4-9}
P(x_1,x_2) \propto q^{x_1+x_2} e^{iu(x_1+x_2)} e^{-v(x_2-x_1)}.
\ee
Using (\ref{4-6}) one obtains for the ``energy'' of this state
\bel{4-10}
\epsilon_u = 2 D \cosh{a} \left(1 - \frac{\cos^2{u}}{\cosh^2{a}}\right).
\ee
Thus the bound state has a non-vanishing energy gap $\Delta \epsilon =
2 D \sinh^2{a}/\cosh{a}$ for any asymmetry $a\neq 0$.

For the determination of the solution of the master equation with initial
condition $\delta_{x_1,y_1}\delta_{x_2,y_2}$ one proceeds in a way analogous
to the totally asymmetric case. First we note that choosing $A_{12}=1$,
multiplying the time-dependent wave function by $\exp{(- ip_1 y_1 - ip_2 y_2)}$
and integrating $p_1$ and $p_2$ from 0 to $2\pi$ gives, at time $t=0$ the
correct initial value $\delta_{x_1,y_1}\delta_{x_2,y_2}$ plus a non-vanishing
term from the reflected wave proportional to $S_{21}$. This term needs to be
cancelled by an appropriate choice of the amplitude of the bound state
contribution. This may be determined by using center of mass coordinates
$R=x_1+x_2$, $R_0=y_1+y_2$ and relative coordinates $r=x_2-x_1>0$,
$r_0=y_2-y_1>0$. Solving first for the $R$-dependence of the master equation
(which is trivial) one obtains then a lattice diffusion equation in a single
coordinate $r$ with partially absorbing boundary condition. This equation
was solved in \cite{Stin95} and one finally finds that
\bea
P(x_1,x_2;t|y_1,y_2;0) & = & \frac{1}{(2\pi)^2} \int_0^{2\pi} dp_1
\int_0^{2\pi} dp_2 e^{-(\epsilon_{p_1} + \epsilon_{p_2})t - ip_1 y_1 - ip_2
y_2}
 \times \nonumber \\
 & & \left( e^{ip_1x_1 + ip_2x_2} + S_{21}
     e^{ip_2x_1 + ip_1x_2}\right)  \nonumber \\
\label{4-11}
 & & + \frac{1}{\pi} \int_0^{\pi} du
\left(e^{2v}-1\right)  q^{R-R_0} e^{iu(R-R_0)}
e^{-v(r+r_0)} e^{-\epsilon_u t} \\
 & \equiv & P^{cont} + P^{bound}
\eea
solves the master equation of the partially asymmetric process with two
particles and initial condition $P(x_1,x_2;t|y_1,y_2;0)= \delta_{x_1,y_1}
\delta_{x_2,y_2}$. The first piece $P^{cont}$ in the sum is the contribution
from the continuum of states (\ref{4-5}), whereas in the second piece
$P^{bound}$ one recognizes the contribution from the bound state. Using the
identity
\bel{4-12}
\mu^{-n} I_n(2\tau) = \frac{1}{2\pi} \int_0^{2\pi} dp
e^{ipn + (\mu e^{ip} + \mu^{-1} e^{-ip})\tau}
\ee
and expanding the denominator of $S_{21}$ in a geometric series in
$z(p_1,p_2) = (D_R e^{-ip_2} + D_L e^{ip_1})/(D_R+D_L)$ one may rewrite
(\ref{4-11}) in terms of a sum of products of two modified Bessel functions.
With (\ref{4-12}) the bound state contribution takes the form
\bel{4-13}
P^{bound} = \frac{1}{\pi} \int_0^{\pi} du e^{iu(R-R_0)}
\left( 1-\xi^2 \right) {\xi}^{r+r_0-2} e^{-( 1-\xi^2 )(D_R+D_L) t}
\ee
where $\xi(u) =  z(u,-u)$.

The general $N$-particle problem is solved
by the Bethe ansatz for $N$ particles and by determining the various
bound state contributions. A way of determining the contributions from
the bound states in the general case is by an appropriate contour integration
in the complex $k_i$ planes which includes the poles of the reflection
coefficients $S_{ij}$. These poles give rise to the bound state contributions
in a general $N$-particle problem.

\section{Diffusion of two particles}
\setcounter{equation}{0}

Here we want to study how the exclusion interaction affects the diffusion
of two particles. In order to get an understanding of what is happening we put
the particles at time $t=0$ on lattice sites $y_1=-1$ and $y_2=1$ and we study
the moments of the density distribution at time $t$. This describes the
diffusive broadening of an initially spatially concentrated density.

We introduce expectation value $\langle n_x \rangle$ which is the probability
of finding a particle on site $x$ at time $t$. The moments of this density
distribution may be obtained from the Fourier transform $\hat{\rho}(q) =
\sum_x e^{-iqx} \langle n_x \rangle$ by taking derivatives w.r.t $q$. Here we
are interested in
\bea
N & = & \hat{\rho}(0) \; = \; 2 \\
\langle X \rangle & = & \frac{i}{N} \hat{\rho}'(0) \; = \;
\frac{1}{2} \sum_x x \langle n_x \rangle\\
\langle X^2 \rangle & = & - \frac{1}{N} \hat{\rho}''(0) \; = \;
\frac{1}{2} \sum_x x^2 \langle n_x \rangle
\eea
from which we shall calculate the asymptotic drift velocity $v$ and the
asymptotic collective diffusion constant $\Delta$ defined by
\bea
\label{5-2a}
v & = & \lim_{t\rightarrow \infty} \frac{d}{dt} \langle X \rangle \\
\label{5-2b}
\Delta & = & \lim_{t\rightarrow \infty} \frac{d}{dt}
\left( \langle X^2 \rangle - \langle X \rangle^2 \right)
\eea

It may be useful to remind the reader of what these quantities are in case
of non-interacting particles. This allows for a comparison of the interacting
and the non-interacting system. For a single particle or for two
non-interacting
particles the density satisfies the diffusion equation
\be
\frac{d}{dt} \langle n_x \rangle = D_R \langle n_{x-1} \rangle +
D_L \langle n_{x+1} \rangle - (D_R+D_L) \langle n_x \rangle.
\ee
Integrating this equation gives after a short calculation
\bea
\label{5-3a}
v & = & D_R - D_L \\
\label{5-3b}
\Delta & = & D_R + D_L
\eea
which is trivial in the sense that this is just a way of defining the driven
non-interacting process. The non-trivial point is the determination of these
quantities for the system with exclusion interaction.

First we note that for the ASEP the density satisfies the continuity
equation
\bel{5-4}
\frac{d}{dt} \langle n_x \rangle = \langle j_{x-1} \rangle -
\langle j_{x} \rangle
\ee
with the current
\bel{5-5}
\langle j_{x} \rangle = D_R \langle n_x (1-n_{x+1}) \rangle -
D_L \langle (1-n_x)n_{x+1} \rangle
\ee

Thus one gets $v=\sum_x \langle j_{x} \rangle/2 = (D_R-D_L) \sum_x
(\langle n_{x} \rangle - \langle n_{x}n_{x+1} \rangle)/2= (D_R-D_L)(1-\sum_x
\langle n_{x}n_{x+1}\rangle/2)$. Using (\ref{4-4b}) one may write
$\sum_x \langle n_{x}n_{x+1}\rangle = \sum_x P(x,x+1;t) =
\sum_x (D_R P(x,x;t) + D_L P(x+1,x+1;t))/(D_R+D_L) = \sum_x P(x,x;t)$.
Thus with the initial condition considered above one gets
\bel{5-6}
\langle X \rangle = (D_R-D_L)\left(t - \frac{1}{2}\int_0^t d\tau
\sum_x P(x,x;\tau|-1,1;0)\right).
\ee

Using the continuity equation the diffusion constant may be written
$\Delta = \sum_x (x+1/2) \langle j_{x} \rangle- 2v\langle X \rangle=
v + 2 D_L + 2 (D_R - D_L -v)\langle X \rangle - (D_R - D_L) \sum_x x
\langle n_{x}n_{x+1}\rangle$. Now (\ref{5-6}) and (\ref{4-4b})
lead to
\bea
\Delta & = &
\lim_{t \rightarrow \infty} \left\{ D_R+D_L + \right.
\nonumber \\
 & & (D_R-D_L)^2  \sum_x P(x,x;t|-1,1;0) \left[ t-\frac{1}{2} \int_0^t d\tau
\sum_x P(x,x;\tau|-1,1;0) \right] \nonumber \\
\label{5-7}
 & &\left. - (D_R - D_L) \left[\sum_x \left(x-\frac{D_R-D_L}{2(D_R+D_L)}\right)
P(x,x;t|-1,1;0) \right] \right\}.
\eea
Therefore in order to determine $v$ and $D$ and has to calculate
$P_0 =\sum_x P(x,x;t|-1,1;0)$ and $P_1=\sum_x x P(x,x;t|-1,1;0)$.
Each of these quantities can be split into three contributions arising
from the two contributions from the continuum and the bound state contribution
in (\ref{4-11}).

There is no contribution from the bound state to $\sum_x P(x,x;t|-1,1;0)$
and defining $\tau = 2(D_R+D_L)t$ one finds
\bel{5-8}
P_0 = e^{-\tau} (I_2(\tau) + I_1(\tau))
\ee
where the term proportional to $I_1$ is the term arising from the reflected
wave proportional to $S_{21}$.

The calculation of $P_1$ is slightly more involved, but still straightforward,
and gives
\bea
P_1 & = & \frac{\tau}{2(D_R+D_L)}e^{-\tau}(D_R I_3 - D_L I_1) +
e^{-\tau}I_2  \nonumber \\
 & & +\frac{\tau}{2(D_R+D_L)}e^{-\tau}(D_R I_2 - D_L I_0) +
\frac{D_R - D_L}{2(D_R+D_L)}(1+e^{-\tau}  I_0) + \frac{D_R}{D_R+D_L}
e^{-\tau} I_1 \nonumber \\
\label{5-9}
 & & - \frac{D_R - D_L}{D_R+D_L}
\eea
where the last piece in the sum comes from the bound state and the arguments
of the Bessel functions are all $\tau$.

Putting everything together and taking the limit $t\rightarrow \infty$
in (\ref{5-7}) finally yields
\bea
\label{5-10}
v & = & D_R-D_L \\
\label{5-11}
\Delta & = & D_R+D_L + \frac{(D_R-D_L)^2}{D_R+D_L}
\left(\frac{1}{2}-\frac{1}{\pi}\right).
\eea
On comparison with the results (\ref{5-3a}), (\ref{5-3b}) for non-interacting
particles
one notices that the exclusion interaction alone does not change the
collective two-particle diffusion constant. In the undriven system one has
$\Delta = D_R+D_L$ as in the non-interacting system. In the presence of
the drift, however, $\Delta$ increases to the value (\ref{5-11}).
This result was obtained for particles placed initially at sites $x_{1,2}=
-1,1$. It is however valid for any (finite) initial separation $r_0$. After
a time $t_0 \gg r_0^2$ the details of the initial condition are washed out.

\section{Conclusions}

The main result of this paper is the solution (\ref{2-17}) of the master
equation for the asymmetric simple exclusion process. This solution allows for
a
complete description of a system of finitely many particles. As a simple
example
we have investigated the collective diffusion of two single particles. We found
that the diffusive broadening of the density profile in the driven system
(\ref{5-11}) is faster than both in the undriven and in the non-interacting
case.

The solution of the master equation may also be used for the analysis of
quantities in systems with finite {\em density}. Consider e.g. a system with
constant non-zero density $\rho$ with a local inhomogeneity such as a lattice
site $y$ where initially the density is $\langle n_y \rangle =1$. In such a
situation it would be interesting to study the time evolution of the density
profile (which gives the dynamical structure function) or the temporal
behaviour
of density correlations. Using the exact solution one can obtain an exact
expansion of these quantities in powers of $\rho$ where the $n^{th}$ power is
obtained by solving
the $n$-particle problem. This can be seen as follows: Suppose one wants to
calculate the time-dependent density profile $\rho_x(t)=\langle n_x (t)
\rangle$
up to second order in the background density $\rho$. The time derivative of the
two-point correlation function involves a three-point correlator which is of
order $\rho^3$ and which therefore may be neglected in the desired second order
approximation. Leaving the three-point correlator out results in a
differential-difference equation for the two-point correlator which is
identical
to the two-particle master equation (\ref{2-1}) with boundary condition
(\ref{2-2}). Thus one can calculate $\langle n_x n_{y} \rangle$ and then by
summing up two-point correlators one gets $\langle n_x \rangle$ up to
order $\rho^2$. For a third order approximation one considers the three-point
correlation function. If one neglects fourth order correlators, it satisfies
the
three-particle master equation. Summing up three-point correlators
yields $\langle n_x \rangle$ up to order $\rho^3$.

A by-product of Sec.~5 is the explicit solution of the two-particle problem for
the symmetric exclusion process.
It is interesting to recall that in the symmetric case $D_R=D_L$ the
conditional probability $P(x_1,x_2;t|y_1,y_2;0)$ determines not only
the behaviour of the two-particle system, but also the behaviour of
various time-dependent density-density correlation functions in $N$-particle
systems, viz.
the equal-time two-point correlator $\langle n_{x_1}(t) n_{x_2}(t) \rangle$
for an arbitrary initial state \cite{Spit70}, the two-time correlator
$\langle n_{x_1}(t_1) n_{x_2}(t_2) \rangle$ for an arbitrary initial state
\cite{Schu94} and the (time-translationally invariant) four-point correlator
$\langle n_{x_1}(t_1) n_{x_2}(t_2) n_{x_3}(t_3) n_{x_4}(t_4) \rangle$
averaged over the stationary distribution \cite{Schu94}. The contribution
of the bound state to pair diffusion was discussed in \cite{Diet83}.
It would be interesting to investigate in detail how the bound state effects
the
behaviour of these correlators in the various space-time regimes of the
symmetric diffusion process.

One puzzling problem is the behaviour of the ASEP in the presence of a
blockage \cite{Wolf90,Tang93,Jano92}, i.e., a bond in the lattice where
particles hop
with rate $r\neq 1$. Numerical and analytical studies seem to indicate that
the steady state current in a finite, half-filled system approaches its maximal
value $j_{max} =1/4$ already at a surprisingly small defect hopping rate
$r \lsim 0.8$ \cite{Jano92,Jano93,Spee96}. This raises the question of a
non-analyticity in the current $j(r)$ for $r<1$ which only an exact
calculation of the steady state current can resolve. A perturbative expansion
in $r$ has been performed up to sixth order using computer algebra
\cite{Jano93}. Each coefficient in the expansion is a rational number with
numerators and denominators rapidly increasing with the order. Many non-trivial
exact steady state properties of exclusion processes have been obtained by
exact
calculation for e.g. small system sizes, then guessing the general structure,
and finally proving that the exact expressions obtained in this way are correct
\cite{SD,DEHP,Derr92,Schu93a,Schu93b,Hinr96}. Unfortunately, applying this
strategy to the perturbative coefficients obtained in \cite{Jano93} seems
hopeless. However, a systematic perturbative expansion of the current may be
performed using the solution (\ref{2-17}) of the master equation where the
$n^{th}$ order in $r$ is obtained by the solution of the $n$-particle problem.
This gives the coefficients of $r^n$ as a sum of $n!$ fractions with slowly
increasing denominators and numerators rather than the single fractions given
in
Ref. \cite{Jano93}. Therefore there is some hope that one might be able to
guess
a pattern in this sequence of fractions. Clearly this is a somewhat
speculative suggestion which very well may turn out to be useless. But it seems
worth trying to obtain exact results in this way.

\section*{Acknowledgments}
The author would like to thank J. Lebowitz, E. Speer and R. Stinchcombe for
stimulating discussions and the Department of Mathematics, Rutgers University,
where part of this work was done, for kind hospitality.

\bibliographystyle{unsrt}

\end{document}